\documentstyle[epsf]{EuroPhys}

\def\etal{{\hbox{{\tenit\ et al.\/}\tenrm :\ }}}

\def\And{{\rm and\ }}

\def\stars{\bigskip\centerline{***}\medskip}

\newif\ifboo \boofalse

\def\Review#1{\boofalse{\it #1},}
\def\Name#1{{\sc #1},}
\def\Vol#1{\ifboo Vol. {\bf #1}\else{\bf #1}\fi}
\def\Year#1{\ifboo #1\else(#1)\fi}
\def\Book#1{\bootrue{\it #1},}
\def\Page#1{\ifboo {\rm p. #1}\else{\rm #1}\fi}

\begin{document}

\euro{}{}{}{}
\Date{}
\shorttitle{G. KORNISS \etal: LONG-RANGE ORDER
IN A QUASI ONE-DIMENSIONAL ETC.}

\title{Long-Range Order in a Quasi One-Dimensional Non-Equilibrium Three-State
Lattice Gas}
\author{G. Korniss\inst{1}, B. Schmittmann\inst{2} \And R.K.P. Zia\inst{2}}
\institute{
\inst{1} Supercomputer Computations Research Institute,
         Florida State University \\
         Tallahassee, FL 32306-4130, USA \\
\inst{2} Department of Physics and
         Center for Stochastic
         Processes in Science and Engineering, \\
         Virginia Polytechnic
         Institute and State University,\\
         Blacksburg, VA. 24061-0435, USA }

\rec{}{}

\pacs{
\Pacs{05}{20$-$y}{Statistical mechanics}
\Pacs{64}{60Cn}{Order-disorder transformations; statistical mechanics of
                model systems}
\Pacs{64}{75$+$g}{Solubility, segregation, and mixing; phase separation}   }

\maketitle

\begin{abstract}
Biased diffusion of two species with conserved dynamics on a $2$$\times$$L$
periodic lattice is studied via Monte Carlo simulations.
In contrast to its simple one-dimensional version on a ring,
this {\em quasi} one-dimensional
model surprisingly exhibits phase separation in its steady state, which is
characterized by one {\em macroscopic} particle cluster. We study the
order parameter and the cluster-size distributions as a function of the system
size $L$, to support the above picture.
\end{abstract}

Driven diffusive lattice gases are among the simplest systems exhibiting
generic non-equilibrium behavior in their steady states \cite{KLS,dds}. Even
the simplest group of these models with purely short-range interactions,
produced by the excluded volume constraint, can display complex phase
diagrams when driven away from equilibrium. The ones with
non-trivial phase diagram typically involve a breaking of translational 
invariance ({\it e.g.}, open boundaries \cite{open}), or more than one species
of particles \cite{HSZ,KSZ}.
The study of multi-species systems is also motivated by fast ionic conductors
with several mobile species \cite{fast_ionic}, water droplets in
microemulsions with distinct charges \cite{water_in_oil}, gel
electrophoresis \cite{gelelectro,Mukamel_gel}, and traffic flow \cite
{traffic}.

One striking feature of the non-equilibrium steady states of some of these
models is that they exhibit spontaneous symmetry breaking or phase
separation in {\em one dimension} (1D), in contrast to equilibrium systems with
short-range interactions, where no long-range order (LRO) exists at finite
temperature, {\it i.e.} with noise present. Indeed, a simple two species
asymmetric exclusion model with {\em open boundaries} display the unusual
phenomenon of spontaneous symmetry breaking in one dimension \cite{EFGM}.
Also, phase separation, analogous to Bose condensation, triggered by defects
has been observed in systems with {\em disorder}, related to simple traffic
flow \cite{Evans}. Recently, a quasi 1D model for the sedimentation of
colloidal crystals \cite{colloid}, and a simple 1D ``cyclic'' three-state
lattice gas \cite{EKKM}, both homogeneous systems with no boundary effects
({\it i.e.}, with ring geometry), have been shown to undergo phase separation.

In this Letter we aim to illustrate that the role of dimensionality
contributing to long-range order and the notion of ``lower critical
dimension'' can be quite subtle in non-equilibrium systems. We focus on the
difference between the 1D and the quasi 1D version of a driven three-state
lattice gas, consisting of holes and two oppositely ``charged'' species of
particles, subject to an ``electric'' field, $E$. {\em Particle-particle
exchanges} (PPE) {\em are allowed} but occur on a much slower time scale than 
the dominant particle-hole exchanges. In two dimensions (2D), 
the system exhibits a
transition from a disordered (homogeneous) to an ordered phase characterized
by a compact strip of particles spanning the system transverse to the field 
\cite{KSZ}. Indeed, even in the {\em absence} of PPE, a phase transition
exists \cite{HSZ}. In one dimension, the behavior of a system without PPE is
clearly trivial. With PPE, it was first believed that the 1D system would
mimic those in higher dimensions and display a transition to an ordered
state. However, an exact solution (with $E=\infty$) was subsequently found
\cite{Sandow}, showing that this system {\em never} orders (though it
exhibits non-trivial cluster size distributions \cite{Sandow} and
microscopic shocks \cite{DJLS}). This behavior seems counter-intuitive,
since blockages, which cause the transition to an ordered state, should be
enhanced in a 1D system. On closer examination, blockages are found to be
``over-enhanced'', in the sense that they occur on microscopic length-scales
and prevent the macroscopic cluster from growing. As a result, the steady
state of the 1D system is characterized by a typical particle cluster size,
which depends only on the microscopics \cite{Sandow} and does not scale with
the system size.

For {\em equilibrium} systems with short-range interactions, LRO cannot exist 
in a 1D chain or in a (quasi 1D) pair of chains. 
In contrast, the behavior of our
model is much more intriguing. Using Monte Carlo simulations and simple
theoretical methods \cite{KSZ}, we investigate the steady states of a 1D
system of $L$ sites and the quasi 1D case of $2$$\times$$L$. Confirming that
there is no LRO in the 1D case, we find that the $2$$\times$$L$ system
behaves as an $L$$\times$$L$ one! In other words, a macroscopic cluster forms
for sufficiently large $E$ and particle density, with density profiles
resembling the mean-field ones closely. For maximal effects, we choose
$E=\infty $, so that we can compare directly with the exactly solvable model
in 1D \cite{Sandow}. We have also considered a range of transverse diffusion
rates, as a further attempt to interpolate between the ordering, $2$$\times$$L$
system and the non-ordering 1D case.

A concise specification of our model consists of a (quasi 1D) periodic
lattice of $2$$\times$$L$ sites. A site ${\bf x}=(x,y)$ can be empty or
occupied by either a positive or a negative particle. Associated with these
occupancies are the standard variables $n_{{\bf x}}^{+}$ , $n_{{\bf x}}^{-}$,
which assume the value $0$ or $1.$ The excluded volume constraint also
implies $n_{{\bf x}}^{+}n_{{\bf x}}^{-}=0$, for any ${\bf x}$. The external
field is chosen so that a positive (negative) particle {\em never} moves in
the $-y$ ($+y$) direction. At each elementary time step, a pair of
neighboring sites (``bond'') is chosen randomly. For an ``$x$-bond'',
particle-hole pairs are exchanged with rate $\Gamma_{\perp}$, while PPE
occurs with rate $\gamma \Gamma_{\perp}$. Similarly, for ``$y$-bonds'',
the rates of the {\em allowed} exchanges are $\Gamma_{\parallel}$ and $\gamma
\Gamma_{\parallel}$, respectively. Note that $\gamma$ sets the time scale for
PPE processes. In the simulations presented here, $\Gamma_{\perp}
=\Gamma_{\parallel}=1$, $\gamma =0.10$. As indicated, we also studied
cases with $\Gamma_{\perp}/\Gamma_{\parallel}<1$, since the $\Gamma
_{\perp}=0$ limit corresponds to two uncoupled 1D systems. Our time unit is
one Monte Carlo step (MCS), during which $4L$ bonds are chosen. We also
restrict ourselves to ``neutral'' systems 
\mbox{$(\sum_{\bf x} n_{\bf x}^{+}=\sum_{\bf x} n_{\bf x}^{-})$} 
at half filling, {\it i.e.}, \mbox{$\overline{m}\equiv 
\frac{1}{2L}\sum_{\bf x}(n_{\bf x}^{+}+n_{\bf x}^{-})=0.5$}. 
On lattices with $L$ ranging
from $10^{2}$ to $10^{4}$, our runs last from $5$$\times$$10^{5}$ to 
$2$$\times$$10^{6}$ MCS. 
Averages, denoted by $\langle \cdots \rangle $, are performed
over the time series, once the system has settled in a steady state. To
ensure that the final state is independent of the initial configuration, we
started the simulations with both disordered (random) and fully ordered
configurations.

\newpage
To characterize the steady state, we define the particle density profile (in 
$y$): 
\begin{equation}
\rho (y)=\frac{1}{2}\sum_{x=1}^{2}(n_{xy}^{+}+n_{xy}^{-})
\end{equation}
and an (unnormalized) order parameter, 
\begin{equation}
Q=\left| \frac{1}{L}\sum_{y=1}^{L}e^{i2\pi y/L}\rho (y)\right| \;,
\label{op}
\end{equation}
Being the magnitude of the lowest Fourier component of $\rho $, $Q$ provides
a sensitive measure of the density inhomogeneities. For the completely
ordered configuration ({\it i.e.}, $\rho =\Theta (\overline{m}L-y) $; 
$\Theta$ being the Heaviside step-function), $Q=\sin (\overline{m}\pi
)/(L\sin (\pi /L))\approx 0.318$ for an infinite system with $\overline{m}
=0.5 $. On the other hand, $\langle Q\rangle $ vanishes in the disordered
phase, up to finite size effects of ${\cal O}(1/\sqrt{L})$. We measured $
j^{+}$, the current of positive particles. By symmetry, the average negative
particle current should be just $-\langle j^{+}\rangle $. Finally, we also
constructed size distributions for both particle and hole clusters by
building histograms with respect to their length along the field in one
fixed ``column'', say for $x=1$. 
\begin{figure}[t]
\vspace{-2.8cm} \hspace*{-1.0cm} 
\begin{minipage}[b]{7cm}
                \hspace*{-0.5cm} \vspace{0.0cm} 
                \epsfxsize=8.2cm \epsfbox{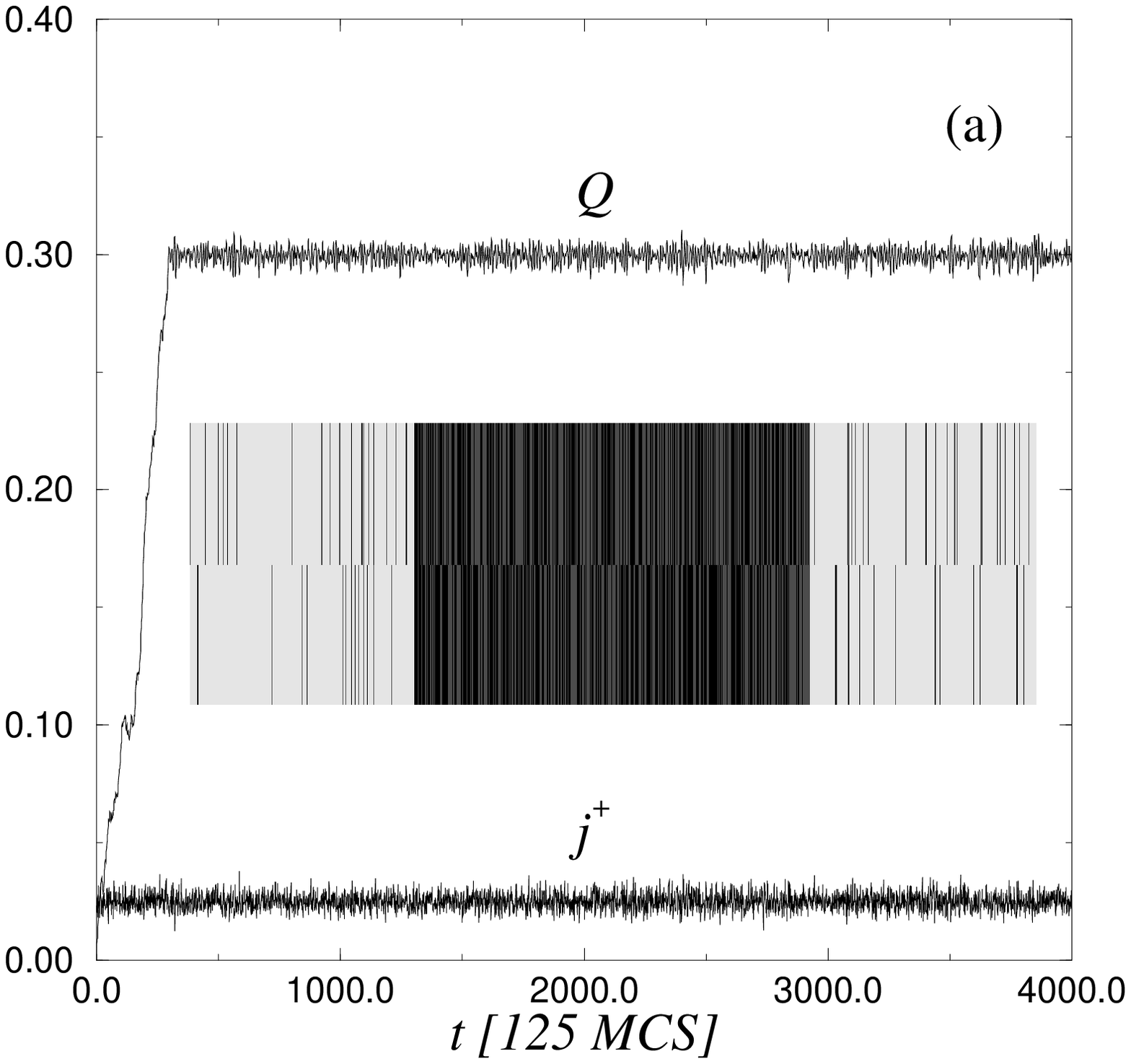} 

\vspace{-1.8cm} 

\end{minipage}
\hspace*{1.0cm} 
\begin{minipage}[b]{7cm}
                \hspace*{0.0cm} \vspace*{0.0cm} 
                \epsfxsize=6.5cm  \epsfbox{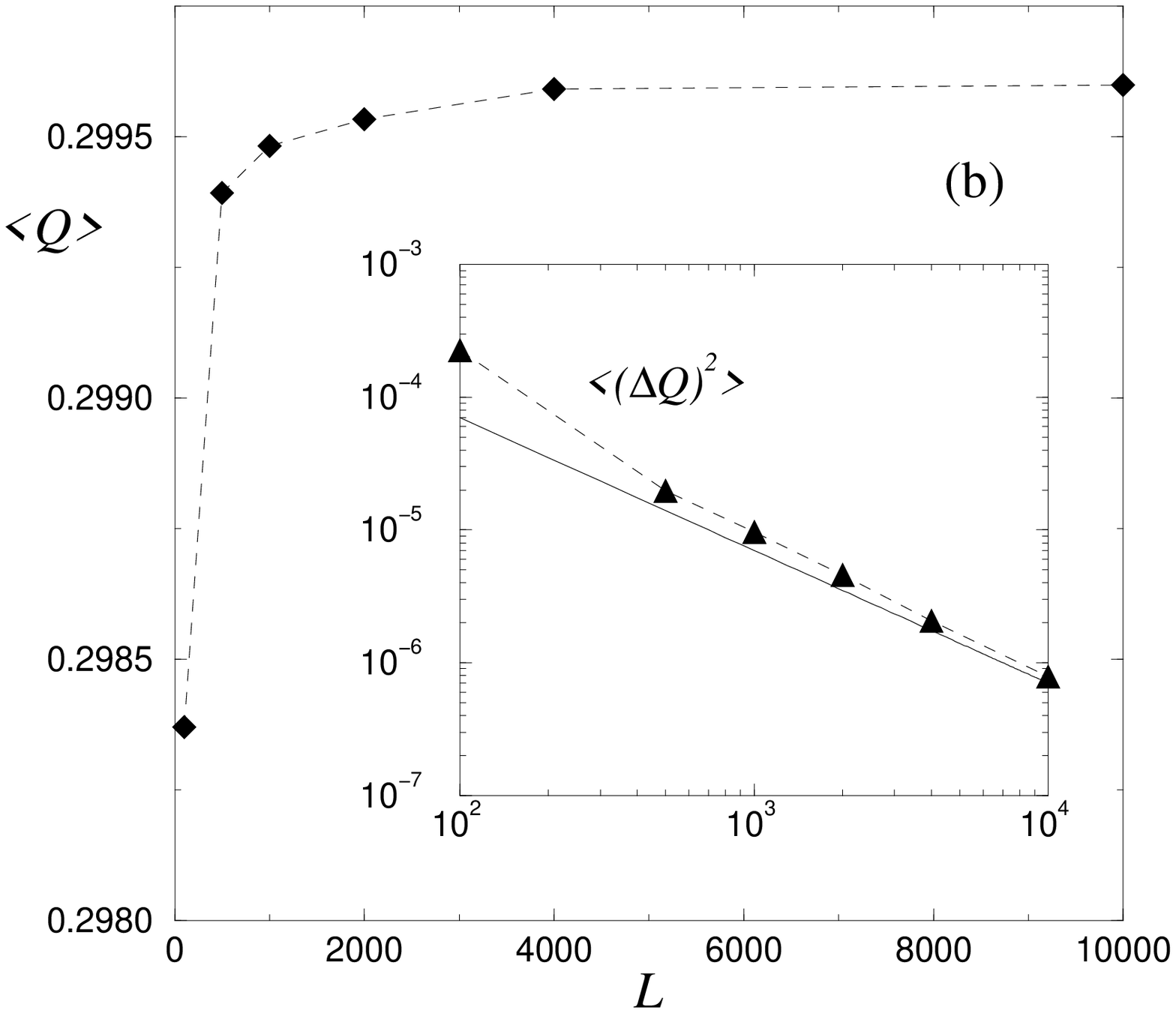}
\end{minipage}
\caption{ (a) Time-series of $Q$ and $j^{+}$ for $L=1000$, starting from
random initial configuration. The inset shows a typical ordered
configuration for the same $L$ (bias points to the right); black, dark gray,
light gray corresponds to $+$'s, $-$'s, and holes respectively. (b)
Finite-size effect for $\langle Q\rangle$ and $\langle (\Delta
Q)^{2}\rangle $ (inset) in the ordered phase. The solid line in the inset
represents a $1/L$ power law. Note the extremely small change in 
$\langle Q\rangle$ (less than $0.5\%$ of its saturation value).}
\end{figure}

For all of our system sizes, we observed that the steady-state configuration
is ordered (inset of fig.\ 1a). Most of the particles ``condense'' into one 
{\em macroscopic} cluster, while the remainder scatters as a small but finite
density of ``travellers'' through the empty region. Due to the infinite
field, holes cannot enter the macroscopic particle cluster. Note that there
will always be a {\em finite} current of either species (proportional to $
\gamma $) in the system, so that coarsening does take place. The growth of
this cluster, when the initial configuration is random, is quite
interesting. After the first $10$-$20$ MCS, small blockages (particle
clusters) form everywhere. After this initial phase, a somewhat slower
process takes over: particle clusters coarsen in time until a single,
macroscopic cluster remains in the system. The evolution of $Q$ (fig.\ 1a)
provides a good picture of this growth process. One important observation is
that the value of the current is basically unaffected after the initial
appearance of small clusters throughout the coarsening process (fig.\ 1a).
This can be understood: the current is mainly controlled by PPE within the
bulk of the particle clusters.

Taken alone, the appearance of a single large cluster is not sufficient for
us to conclude the presence of an ordered state. Indeed, even for the 1D
case, in which there is a finite typical particle cluster size ($
l_{p}^{*}\sim 4\gamma ^{-2}$ for $\overline{m}=0.5$) \cite{Sandow}, we would 
generally observe a single cluster if $L\ll $ $l_{p}^{*}$. 
Although the range of our $L$'s does
exceed $4\gamma ^{-2}=400$, we probe more deeply by analyzing the $L$
dependence of various quantities. The conclusion is that, if $\gamma
L^{\omega} \stackrel{>}{\sim} 1$ were the criterion for the $2$$\times$$L$ 
system to display 1D behavior, 
then $\omega \simeq 0.$ In other words, we believe that
the collective properties of the quasi 1D system are distinct from the
non-ordering 1D chain, but fall within the class of the higher dimensional
systems. In the remainder of this letter, some details of our analyses are
provided.

In fig.\ 1b, we show the average order parameter, $\langle Q\rangle $, as a
function of $L$. It saturates rapidly; even with $L=100,$ it is within 0.5\%
of the apparent $L=\infty $ limit. Meanwhile, its fluctuations, $\langle
(\Delta Q)^{2}\rangle =\langle Q^{2}\rangle -\langle Q\rangle ^{2}$, decay
like $1/L$ (inset fig.\ 1b). The distribution of the sizes of {\em particle}
clusters reveals more of the nature of the steady state. Denoting by $P(l,L)$
the probability of finding a cluster of length $l$ in a system of size $L$,
we are not surprised that there are two well-separated components, one
corresponding to the single macroscopic cluster and the other to the small
clusters of ``travellers''. The macroscopic component appears as a peak at $
l_{0}$. We found that $l_{0}\approx 0.471L$, which is a result of 
$\overline{m}=0.5$ and a clear signal of the large cluster 
scaling linearly with the system size. 
Meanwhile, $\delta l_{0}$, the standard deviation, grows as $
\sqrt{L}$. Focusing on this component alone, we renormalize its integral to
unity and rewrite it in terms of $u\equiv (l-l_{0})/\delta l_{0}$. The
result, $p(u,L)$, is well fitted by a Gaussian (fig.\ 2a). Deviations from
the Gaussian, seen especially for small $L$, carry non-trivial
information on finite size effects which should be investigated further. 
In the ``travellers'' component, the
distribution decays exponentially ($\propto e^{-l/\lambda }$), with $\lambda
<1$ and {\em independent} of $L$. Similarly, the size distribution of the 
{\em hole} clusters is also a simple exponential, independent of $L$ apart
from finite size effects. Both of these exponentials can be understood \cite
{irina} from the observation that the small density of ``travellers'' is
roughly homogeneous. Following approaches to the percolation problem 
\cite{perc}, we consider another interesting distribution, namely, the 
probability that a
randomly selected particle belongs to a cluster of length $l$ (at a given
time). Clearly proportional to $lP(l,L)$, it will be referred to as the
``residence'' distribution. To compare distributions from different $L$'s,
we define $w\equiv l/(\overline{m}L)\in (0,1]$. Thus, $\tilde{P}(w,L)$ is
the probability (density) that a particle ``resides'' in a cluster of length 
$w\overline{m}L$. Since there are essentially no holes in the particle
clusters, $w$ is just the fraction of all the particles in such a cluster.
In fig.\ 2b, we see that $\tilde{P}(w,L)$ peaks at approximately $0$ and at  
$w_{0}\approx 0.942$.
The areas under each component are essentially independent of system size,
while both sharpen with increasing $L$. With finite $L$, the former peaks at 
$1/(\overline{m}L)$, with area $0.058.$ Of course, the complement area is
under the other component. It is not a coincidence that this area ($0.942$)
is identical to the peak position, $w_{0}$, since both are simply 
related to the fraction of particles in the macroscopic cluster. Finally, 
note that this quantity is also $l_{0}/(\overline{m}L).$ 
\begin{figure}[t]
\vspace*{0.0cm} \hspace*{-0.2cm} 
\begin{minipage}[b]{5cm}
                \hspace*{0.0cm} \vspace{0.0cm} 
                \epsfxsize=6.5cm \epsfbox{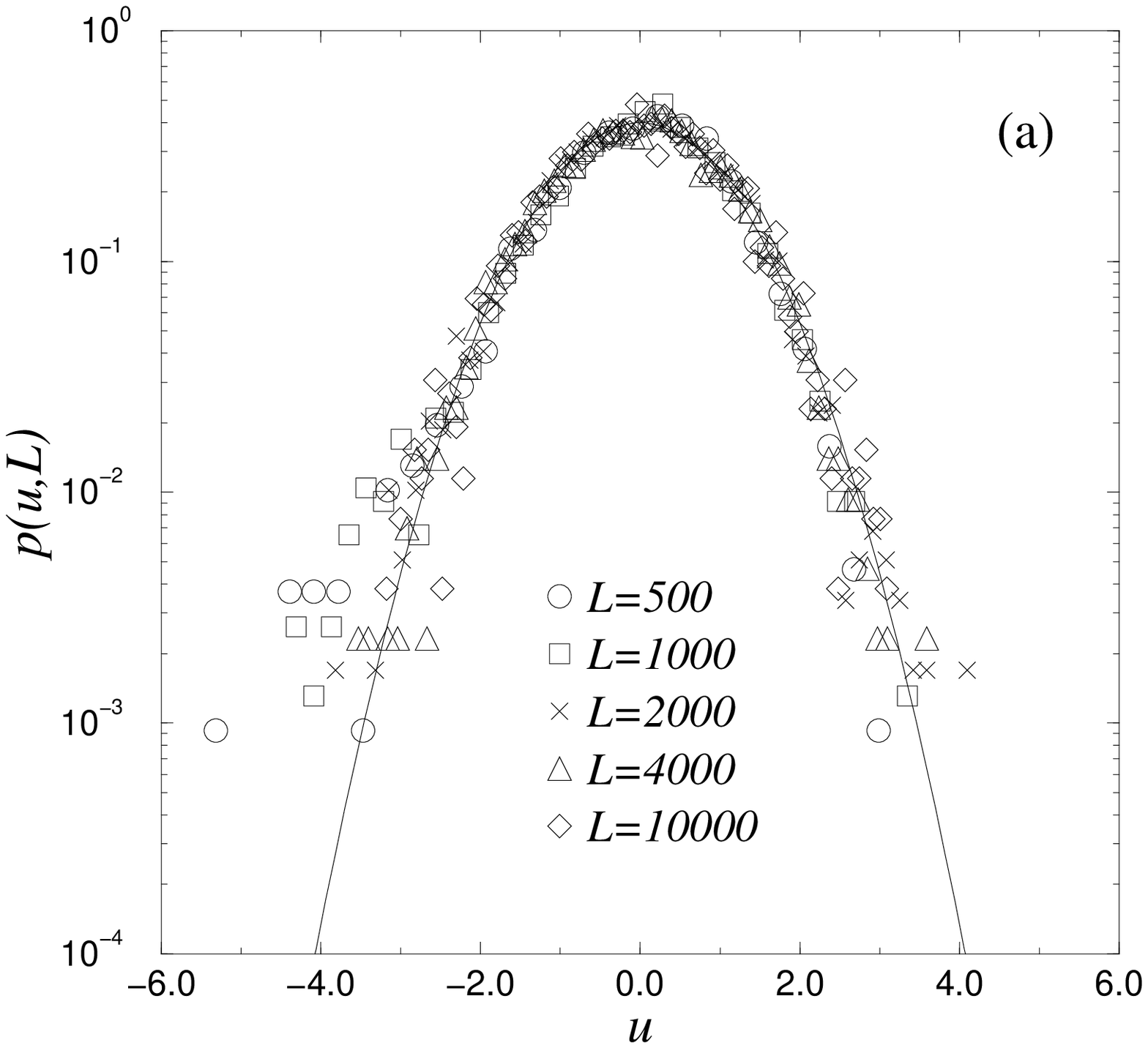} 
\end{minipage}
\hspace*{2.2cm} 
\begin{minipage}[b]{5cm}
                \hspace*{0.0cm} \vspace*{0.0cm} 
                \epsfxsize=6.5cm  \epsfbox{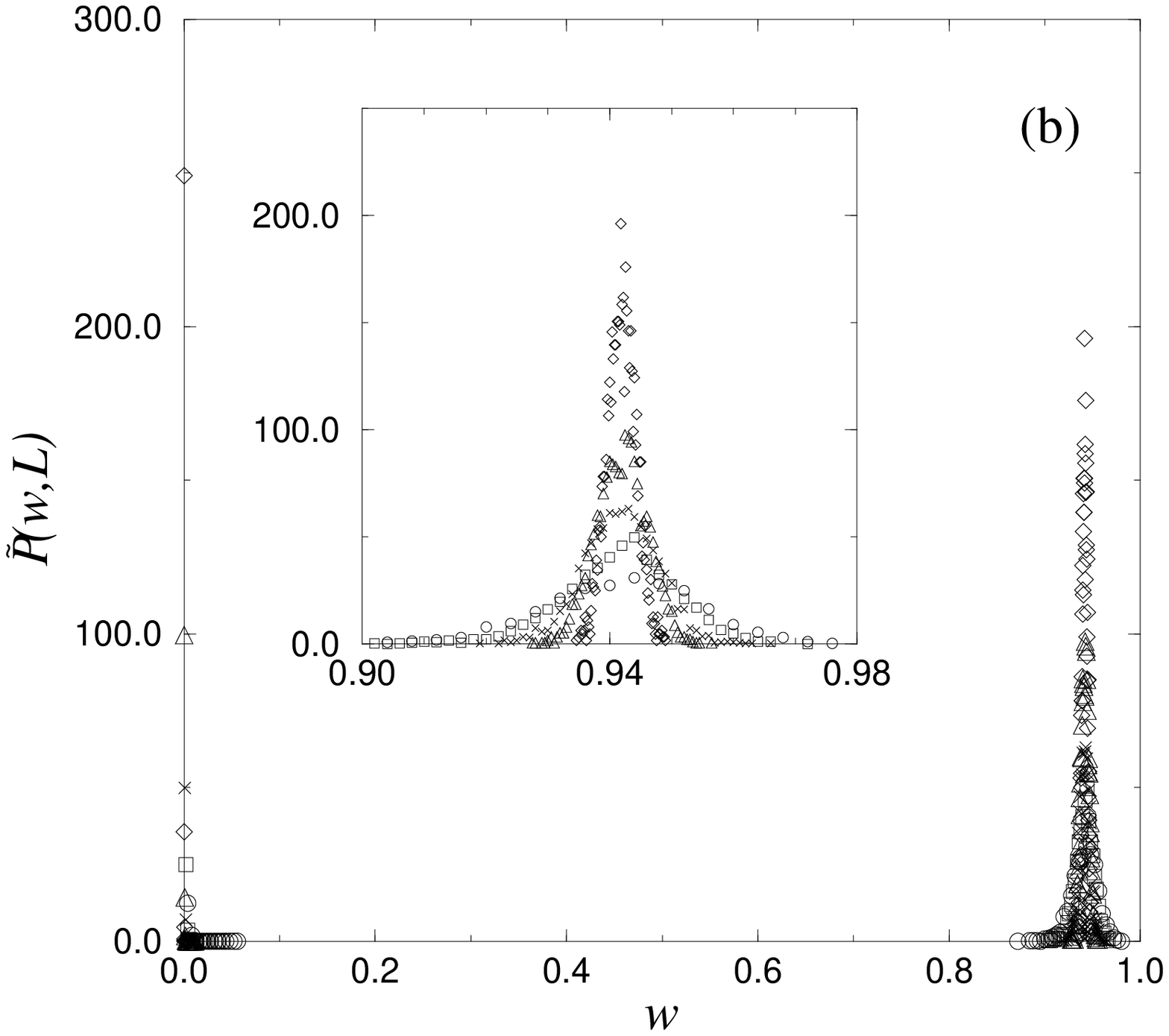}
\end{minipage}
\caption{ (a) The distribution of the relative size-fluctuations
of the macroscopic particle clusters for different system sizes. 
The solid line is a Gaussian with zero mean and
unit standard deviation. (b) The scaled residence distributions. System
sizes and symbols are the same as in (a). The inset gives a closer view of
the distribution around $w_{0}=0.942$.}
\end{figure}

We also studied the system with varying transverse hopping rates:
$0 \leq \Gamma_{\perp}/\Gamma_{\parallel}\leq 1$. Our results indicate that a 
crossover to the 1D behavior occurs at a very small but {\em finite} value of 
this ratio (of the order of $10^{-3}$); only a very careful finite-size 
analysis could reveal whether this occurs when the transverse hopping rate 
becomes comparable to the inverse relaxation time of the 1D system.

Most of our data can be understood by simple mean-field considerations \cite
{KSZ}. As we already pointed out, the current of one species ({\it e.g.},
that of the $+$'s) is controlled by the bulk of the particle cluster.
Although there is a non-trivial charge distribution within the particle
cluster \cite{irina}, it is fairly flat (up to finite size effects) around
the {\em center} of the cluster: thus at the simplest level we can
approximate the density of both species there to be equal to $1/2$, and $
\langle j^{+}\rangle \approx \gamma /4$. Since the steady state is
stationary, the current must be constant and homogeneous through the system,
so that $\langle j^{+}\rangle \approx \gamma /4$ outside the particle
cluster as well. Assuming a homogeneous density of ``travellers'', $m^{*}$,
outside the macroscopic particle cluster, we have \cite{KSZ} 
\mbox{$\gamma/4=(m^{*}/2)[(1-m^{*})+\gamma(m^{*}/2)]$}, yielding $
m^{*}=\gamma /(2-\gamma )$. For $\gamma =0.1$, $m^{*}=0.0526$, which is
within $5\%$ of the Monte Carlo value. The mean macroscopic cluster size 
$l_0$ can also be deduced, since \mbox{$l_{0}+m^{*}(L-l_{0})=\overline{m}L$}. 
The result, $0.472L,$ is very close to the one obtained from the measured
particle cluster size distributions. Also, the order parameter for such an
ordered configuration is just $Q=0.30$, which compares well with the data
(fig.\ 1a). Since the velocity of a tagged particle is approximately $\gamma
/2$ in the particle cluster and \mbox{$(1-m^{*})+\gamma(m^{*}/2)$} outside
the cluster, it is easy to show that the typical time it spends diffusing
through the macroscopic cluster is a fraction $l_0/(\overline{m}L)$ of the
total time needed to travel through the system once. Thus, for our
parameters, a randomly selected particle is found with probability $0.944$
in the large particle cluster, which reproduces the location of the second
peak of the distribution $\tilde{P}(w,L)$ quite well.

In summary, we have studied the non-equilibrium steady state of a driven,
quasi one-dimensional, three-state stochastic lattice gas, using simulations
and simple mean-field arguments. We found that, for the parameter regime
studied, the properties of a {\em coupled pair }of chains of length $L$ are
drastically different from those of the strictly 1D system, {\it i.e.,} the
single chain. With as little as $1\%$ of cross-chain (transverse) moves, the 
$2$$\times$$L$ model already develops a single macroscopic blockage, similar to
the 2D ($L$$\times$$L$) system. Moreover, the finite-size analysis of the
ordered phase, up to $L=10^{4}$, shows no indication of a crossover to 1D
behavior. Thus, we conjecture that LRO survives the $L\rightarrow \infty $
limit in our simple model (fixed $\gamma $ and $\Gamma _{\perp }=\Gamma
_{\parallel }$). If this conjecture proves to be true, our model would pose
a stark contrast to, {\it e.g.,} the $M$$\times$$L$ equilibrium Ising model
where LRO does not exist in the fixed $M$, $L\rightarrow \infty $ limit.
Instead, the strictly one-dimensional system is a ``singular limit'', in
much the same way that entirely different behaviors are found, obviously, in 
$1$$\times$$L$ vs. $2$$\times$$L$ systems of biased diffusion of two species 
{\em without} particle-particle exchange in finite field. The key difference 
between the strictly and the quasi 1D systems appears to be
the ease with which particles can pass one another, leading to the formation
of a macroscopic cluster. Work is in progress to 
reveal further differences between the 1D and the quasi 1D systems, {\it e.g.},
details of the coarsening process leading to LRO \cite{irina}.

\section{Note Added}
After submitting our manuscript we became aware of the work by Arndt 
{\it et al.} \cite{Rittenberg}. They studied a 1D model on a ring, similar to 
ours, but with different exchange rates.

\stars

Stimulating discussions with S. Sandow, C. Godr\`{e}che and D. Mukamel are
gratefully acknowledged. Special thanks to Z. Toroczkai for his comments 
on the manuscript.
We thank E. Ben-Naim and G. Lythe for their support through the DOE at the 
CNLS Nonequilibrium Dynamics Workshop (April 1998) where this work was first 
presented. G. Korniss also wishes to thank P. A. Rikvold and M. A. Novotny 
for their support during the preparation of this work and their critical
reading of the manuscript. This research at Virginia Tech was supported in 
part by NSF Grant No. DMR-9727574.
Work at the Florida State University was supported by SCRI under the US DOE
Contract No. DE-FC05-85ER25000, and by NSF Grant No. DMR-9520325.


\begin{thebibliography}{99}
\bibitem{KLS}  \Name{Katz S., Lebowitz J. L. \And Spohn H.} \Review{Phys.
Rev. B.} \Vol{28} \Year{1983} \Page{1655}; \Review{J. Stat. Phys.} \Vol{34}
\Year{1984} \Page{497}.

\bibitem{dds}  \Name{Schmittmann B. Zia R. K. P.} in \Book{Phase Transitions
and Critical Phenomena} edited by \Name{Domb C. \And Lebowitz J. L.} \Vol{17}
(Academic Press, N.Y.) \Year{1995}.

\bibitem{open} \Name{Krug J.} \Review{Phys.Rev. Lett.} \Vol{67}
\Year{1991} \Page{1882}; 
\Name{Sch\"{u}tz G. \And Domany E.} \Review{J. Stat. Phys.} \Vol{72} 
\Year{1993} \Page{277};
\Name{Derrida B., Evans M. R., Hakim V. \And Pasquier V.}
\Review{J. Phys. A} \Vol{26} \Year{1993} \Page{1493}.

\bibitem{HSZ}  \Name{Schmittmann B., Hwang K. \And Zia R.K.P.} 
\Review{Europhys. Lett.} \Vol{19} \Year{1992} \Page{19}.

\bibitem{KSZ}  \Name{Korniss G., Schmittmann B. \And Zia R. K. P.}
\Review{Europhys. Lett.} \Vol{32} \Year{1995} \Page{49}; \Review{J. Stat.
Phys.} \Vol{86} \Year{1997} \Page{721}.

\bibitem{fast_ionic}  See {\it e.g.}, \Name{Chandra S.} \Book{Superionic
Solids. Principles and Applications} (North Holland, Amsterdam) \Year{1981}.

\bibitem{water_in_oil}  \Name{Aertsens M. \And Naudts J.} \Review{J. Stat.
Phys.} \Vol{62} \Year{1990} \Page{609}.

\bibitem{gelelectro}  \Name{Rubinstein M.} \Review{Phys.Rev. Lett.} \Vol{59}
\Year{1987} \Page{1946}; \Name{Duke T. A. J.} \Review{Phys. Rev. Lett.}
\Vol{62} \Year{1989} \Page{2877}; \Name{Schnidman Y.} in \Book{Mathematics
in Industrial Problems IV} edited by \Name{Friedman A.} (Springer, Berlin)
\Year{1991}; \Name{Widom B., Viovy J. L. \And Desfontaines A. D.} \Review{J.
Phys I} \Vol{1} \Year{1991} \Page{1759}.

\bibitem{Mukamel_gel}  \Name{Alon U. \And Mukamel D.} \Review{Phys. Rev. E}
\Vol{55} \Year{1997} \Page{1783}.

\bibitem{traffic}  \Name{Biham O., Middleton A. A. \And Levine D.} 
\Review{Phys. Rev. A} \Vol{46} \Year{1992} \Page{R6128}; \Name{Leung K.-t.}
\Review{Phys. Rev. Lett.} \Vol{73} \Year{1994} \Page{2386}.

\bibitem{EFGM}  \Name{Evans M. R., Foster D. P., Godr\`{e}che C. \And
Mukamel D.} \Review{Phys. Rev. Lett.} \Vol{78} \Year{1995} \Page{208};
\Review{J. Stat. Phys.} \Vol{80} \Year{1995} \Page{69}.

\bibitem{Evans}  \Name{Evans M. R.} \Review{Europhys. Lett.} \Vol{36}
\Year{1996} \Page{13}; \Name{Krug j. \And Ferrari P. A.} \Review{J. Phys. A.}
\Vol{29} \Year{1996} \Page{L213}.

\bibitem{colloid}  \Name{Lahiri R. \And Ramaswamy S.} \Review{Phys. Rev.
Lett.} \Vol{79} \Year{1997} \Page{1150}.

\bibitem{EKKM}  \Name{Evans M. R., Kafri Y., Koduvely H. M. \And Mukamel D.}
\Review{Phys. Rev. Lett.} \Vol{80} \Year{1998} \Page{425}; cond-mat/9802255
Preprint, 1998; see also \Name{Arndt P. F., Heinzel T. \And 
Rittenberg V.} \Review{J. Phys. A.} \Vol{31} \Year{1998} \Page{L45}.

\bibitem{Sandow}  \Name{C. Godr\`{e}che \And S. Sandow} (private
communications, to be submitted).

\bibitem{DJLS}  \Name{Derrida B., Janowsky S. A., Lebowitz J. L. \And Speer
E. R.} \Review{Europhys. Lett.} \Vol{22} \Year{1993} \Page{651}; \Review{J.
Stat. Phys.} \Vol{73} \Year{1993} \Page{813}.

\bibitem{irina}  \Name{Mazilu I., Korniss G. \And Schmittmann B.} (in
preparation).

\bibitem{perc}  See {\it e.g.}, \Name{Stauffer D.} \Book{Introduction to
Percolation Theory} (Taylor \& Francis, London and Philadelphia) \Year{1985}.

\bibitem{Rittenberg} \Name{Arndt P. F., Heinzel T. \And Rittenberg V.}
cond-mat/9809123 Preprint, 1998.


\end{thebibliography}
\end{document}